\documentclass[onecolumn,pra,a4paper,showpacs,floatfix,preprintnumbers,amsmath,amssymb]{revtex4}

\usepackage{graphicx}
\usepackage{dcolumn}
\usepackage{bm}

\newcommand{\ud}{\mathrm{d}}

\newcommand{\I}{\mathrm{I}}
\newcommand{\II}{\mathrm{II}}

\newcommand{\be}{\begin{equation}}
\newcommand{\ee}{\end{equation}}

\begin{document}

\title{Dephasing and breakdown of adiabaticity in the splitting of Bose Einstein
Condensates}

\author{L. Pezz\'e$^{1,2}$, 
A. Smerzi$^{1,2}$, G.P. Berman$^1$, A.R. Bishop$^1$ and L.A. Collins$^1$}
\affiliation{1) Theoretical Division, Los Alamos National Laboratory,
Los Alamos, New Mexico 87545, USA\\
2) Istituto Nazionale per la Fisica della Materia BEC-CRS\\
and Dipartimento di Fisica, Universit\`a di Trento, I-38050 Povo, Italy}

\date{\today}

\begin{abstract} We study the quantum dynamics of a
BEC condensate trapped in a double-well potential with a rising 
interwell barrier.
We analytically find the characteristic time scales of the splitting
process and compare our results with numerical analyses available
in the literature. In first stage of the dynamics, the condensate follows
adiabatically the rising of the interwell barrier. At a critical time 
$t_{ad}$, 
small amplitude fluctuations around the average trajectory
increase exponentially fast, signaling the break-down of adiabaticity.
We have found a highly non-trivial dependence of the
dephasing time $t_D$, defined by 
$\sigma_{\phi}(t_D)=1$, where $\sigma_{\phi}(t)$ is the
dynamical quantum phase spreading, on $t_{ad}$ and on the ramping time
of the interwell barrier.\\
\end{abstract}

\maketitle

\emph{Introduction: }From the early observations of a Bose
Einstein Condensate (BEC) there has been a growing interest in
theoretical and experimental studies of a condensate in a
double-well trap. There are several goals related to these
studies: i) to understand the analog of the Josephson effect in this
type of system; ii) to clarify the meaning of the phase in
quantum mechanics; and iii) to create interferometers working at the
Heisenberg limit. The recent experimental realization
\cite{Shin_2004} of a stable double-well trap is renewing the interest
in these topics. Even though residual sources of noise are still
limiting the coherence life time of the two coupled systems, 
the experimental progress is very promising \cite{Lee_2004} for
future developments and possible technological applications.

In this paper we study, in a two-mode approximation, the dynamical 
splitting of a condensate trapped in a 
double-well while ramping up the interwell barrier.
This problem has raised some controversy in the past.
In \cite{Javanainen_1997}, Javanainen and Wilkens (JW) analyzed the
condensate splitting in a two-stage stages model: first, an initial
condensate is partially split in two parts by ramping a potential
barrier in the middle. The ramping rate is $R_{\I}\ll \omega_p$,
where $\omega_p$ is the Josephson plasma frequency. Because of
this inequality the process is strictly adiabatic. It ends in a
regime when it is possible to neglect the particle exchange
between the two wells. In the second stage, the barrier is rised
to infinity at a rate $R_{\II}\gg \omega_p$, and then the system is
left alone for a time $t$. There has been a debate between JW and
Legget and Sols (LS) \cite{Legget_Sols_1998} about the rate at
which the two halves of the condensate lose their phase memory in
this second stage. The crucial point was the estimation of the
ground state phase fluctuations. In \cite{Javanainen_1999},
Javanainen and Ivanov (JI) clarified the issue by analyzing numerically
a two-mode model of a realistic continuous splitting.

In the present paper, we study the splitting
problem within an analytically solvable 
model. We recover simple analytical estimates of the various time scales
of the problem, which agree quite well with the JI estimates.
Moreover, we emphasize that even after the loss of adiabaticity
it is still not possible to neglect the tunneling exchange between the
wells. As a consequence, 
the dephasing time, in which the system loses memory about the relative 
phase between the two condensates, has a complicated dependence on the
ramping time of the interwell barrier.
This has important consequences when studying the realizability of a 
BEC interferometer.

The main results of this paper are the analytical expression of 
the time of breakdown of adiabaticity and the dephasing time as a function of  
both the initial condition and the typical time scale of the splitting process.\\

\emph{The Quantum-Phase model. } In this section, we review the
Quantum-Phase model (QPM)
of a condensate trapped in a symmetric two-well trap at zero
temperature.
The second
quantization Hamiltonian of a system of bosons interacting with a
$\delta$-pseudopotential is given by 
\be 
\hat{H}(t)= \int \ud z \,
\hat{\Psi}^{\dag}(z,t) \,
\bigg(-\frac{\hbar^2}{2m}\frac{\partial^2}{\partial z^2}+V(z,t)
\bigg) \, \hat{\Psi}(z,t)+ \frac{g}{2}\int \ud z \,
\hat{\Psi}^{\dag}(z,t) \hat{\Psi}^{\dag}(z,t) \hat{\Psi}(z,t)
\hat{\Psi}(z,t), 
\ee 
where $\hat{\Psi}$ is the bosonic field
operator, and $V(z,t)$ is the time-dependent external double-well
potential; $g= {{4 \pi \hbar^2 a}\over m}$ 
is the strength of the interparticle interaction, with 
$a$ being the $s$-wave scattering length. The two-mode ansatz reads
\be \label{ansatz}
\hat{\Psi}(z,t)=\psi_1(z,t)\, \hat{a}_1 + \psi_2(z,t)\, \hat{a}_2, 
\ee
where $\psi_{1,2}(z,t)$ can be constructed as sum and difference of the 
first symmetric and antisymmetric Gross-Pitaevskii dynamical
wave-functions in the double well trap.
The operator $\hat{a}^{\dag}_{1,2}$ ($\hat{a}_{1,2}$)
creates (destroys) a particle in the modes 1,2, respectively. In
the following, we will decouple the ``external dynamics'' (the
evolution of the wave function) and the ``internal dynamics'' (the
evolution of the operators) \cite{Menotti_2001}. The important
time scale of the external dynamics is given by the trapping
oscillation period $\tau_z=2\pi/\omega_z$, where $\omega_z$ is the 
trap frequency. If the barrier is
rised on a time scale $\Delta t \gg \tau_z$, then the final wave
function $\psi_{1,2}(z,t)$ will correspond to the ground state of
the 1,2 well respectively. If $\Delta t \ll \tau_z$, on the other hand,
the splitting process
excites the system, increasing the condensate energy. 
Menotti \emph{et al.} in \cite{Menotti_2001} indicate 
the revival time of coherence,$\tau_r$ \cite{Lewenstein_1996,Villain_1997},
as an upper bound for the internal dynamics. 
In fact, when $\Delta t \gg \tau_r$ the phase
coherence is lost during the splitting process, and the two final
condensates will be independently exhibiting no phase coherence.
In current experiments, $\tau_z \ll \tau_r$, with
$\tau_r$ longer than the life time of the condensate. Therefore, raising
the potential barrier at a rate $\tau_z \ll \Delta t \ll \tau_r$,
we can decouple the internal and the external dynamics. In the two-mode 
approximation the Hamiltonian of the system is
\cite{Javanainen_1999, Anglin_2001} 
\be \label{HBH}
\hat{H}=\frac{E_c}{4}\Big(\hat{a}^{\dag}_1\hat{a}^{\dag}_1\hat{a}_1\hat{a}_1+
\hat{a}^{\dag}_2\hat{a}^{\dag}_2\hat{a}_2\hat{a}_2\Big)
-\frac{E_j}{N}\Big(\hat{a}^{\dag}_1\hat{a}_2+\hat{a}^{\dag}_2\hat{a}_1
\Big). 
\ee 
The operator
$\hat{N}=\hat{n}_1+\hat{n}_2=\hat{a}^{\dag}_{1}\hat{a}_{1}+\hat{a}^{\dag}_{2}\hat{a}_{2}$
is the total number of particles, and it commutes with $\hat{H}$.
The quantity $E_j$ is the ``Josephson coupling energy'', and $E_c$
is the ``one-site energy'': 
\be 
E_c=2 g \int \ud z \,|\psi_{1}(z,t)|^2=2 g \int \ud z \, |\psi_{2}(z,t)|^2, 
\ee
\be E_j=-N
\int \ud z 
\bigg[\frac{\hbar^2}{2m}\frac{\partial \psi_1(z,t)}{\partial
z}\frac{\partial \psi_2(z,t)}{\partial z} + \psi_1(z,t) V(z,t)
\psi_2(z,t) \bigg]. 
\ee 
It is convenient to study Eq.(\ref{HBH}) in the Bargmann 
phase-states representation \cite{Anglin_2001}. We write a general state
in the Hilbert space of the two-mode system as
\be \label{SB}
|\psi\rangle=\int_{-\pi}^{+\pi}\frac{\ud \phi}{2\pi} \, \psi(\phi) \, |\phi
\rangle,
\ee
where $\phi$ is the relative phase between the two modes, and 
\be \label{BV}
|\phi \rangle=\sum_{n=-N/2}^{N/2} \frac{e^{i n \phi}}{\sqrt{n!}} |n \rangle 
\ee
are un-normalized vectors of the overcomplete Bargmann base, written 
in the relative number of 
particles $n$. In the Bargmann representation the action of any operators on 
$|\psi\rangle$ can be represented in terms of differential operators acting 
on the associated $\psi(\phi)$. The main consequence of the overcompleteness is
the non-standard inner product between Bargmann vectors (\ref{BV}) 
$\langle \, \phi \, | \, \theta \, \rangle \approx
\cos^N\big(\frac{\phi-\theta}{2}\big)$.
It affects the inner product between
states (\ref{SB})  written in the Bargmann representation. \\
In the limit
$E_j \ll N^2 E_c$,
and rescaling the time as $t \to t\, E_c/2\hbar$, 
the dynamical equation for the $2 \pi$-periodic phase amplitude
$\Psi(\phi,t)$ is 
\be \label{dynamical_equation}
i \frac{\partial \Psi(\phi, t)}{\partial t}= - \frac{\partial^2
\Psi(\phi, t)}{\partial \phi^2} - \Gamma(t) \cos(\phi) \Psi(\phi,
t), \ee
with $\Gamma(t)=2 E_j(t)/E_c$.
In this paper, we study the dynamical evolution of the wave function
$\Psi(\phi, t)$ when the initial condensate is split
with a symmetric double-well potential. This experiment has been recently
realized \cite{Shin_2004}, with the two wells ramped apart linearly in time. The
distance between the center of the wells evolves according to
$d(t)=d_0+d_{fin}t/\Delta t_R$, where $d_0$ and $d_{fin}$ are the
initial and final distances, respectively, and $\Delta t_R$ is the
total ramping time. With this setup it is possible to find, both
in WKB approximation \cite{Messiah} and by a numerical 1-D simulation, that
the Josephson coupling energy evolves in time with the exponential
law $E_j(t)=E_j(0)\,e^{-t/\tau}$, where the effective ramping time
$\tau=\Delta t_R \, \hbar/\sqrt{2m(V_0-\mu)d_0^2}$ depends of the
particle mass $m$, the initial height of the potential barrier
$V_0$, and the chemical potential $\mu$.
In the two-mode approximation the one-site energy $E_c$ remains 
constant during the dynamics. We notice that 
$Ej$ scales exponentially with the interwell distance only
when the condensates are well separated. 
During the initial splitting, when the chemical potential is close to the
interwell barrier, the dynamics remains adiabatic. 
The adiabaticity
will break down at a large separation of the two condensates, in the deep 
tunneling regime, which will be the focus of the next sections.\\

The two-mode model has been extensively discussed in the literature.
In general, it is expected to work when the ground state and the
first excited state are close to each other in energy and well
separated from higher energy modes. It works in the limit of weak
atom-atom interaction \cite{Javanainen_1999} and small atom
number, $N g \ll \omega_z$, where $\omega_z$ is the frequency of the
trap in the separation direction. It also has limited validity in
the case of a low potential barrier when it is not allowed to
neglect higher excitation modes. However, it becomes increasingly
accurate by rising the potential barrier. In this case, in fact,
the two lower lying modes become closer 
in energy and separated from the higher ones. As pointed 
out by Menotti \emph{et al.} \cite{Menotti_2001}, a fundamental 
condition for the validity of the two-mode model is the different time scale 
which characterizes the ``internal'' and ``external'' dynamics. This decoupling 
is at the basis of the two-mode ansatz (\ref{ansatz}). Through a gaussian 
variational model, and numerical simulations of the mode functions \cite{Menotti_2001}, 
it is possible to conclude that, to a good approximation and for reasonable regimes, the 
wave function dynamics are independent of the operator dynamics. \\

\begin{figure}[!h]
\begin{center}
\includegraphics[scale=0.8]{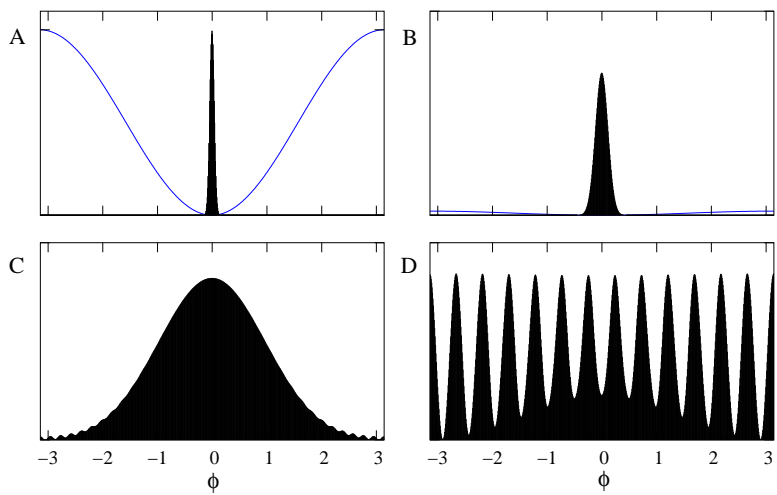}
\end{center}
\caption{\small{Profile of the phase amplitude $|\Psi(\phi)|^2$ at
different times: A) $t=0$, corresponds to the ground state at the
beginning of the dynamic; B) $t=t_{ad}=23$ msec, which corresponds to
the breakdown of adiabaticity as calculated by Eq. (\ref{t_ad});
C) $t=t_D=200$ msec, corresponds to the dephasing time
($\sigma_{\phi}(t_D)=1$) as calculated by the equation (\ref{t_D}); D)
$t=500$ msec. The blue line represent the cosine potential (in
rescaled units) decreasing with $\tau=5$ msec. }}\label{wave}
\end{figure}

\emph{Numerical Solution. } We first solved numerically equation
(\ref{dynamical_equation}). Figure (\ref{wave}) presents a
plot of the normalized $|\Psi(\phi, t)|^2$ at different times.
Superimposed on the phase amplitude, the blue line presents the
cosine potential in arbitrary units. At $t=0$ (Fig. (\ref{wave},A))
the phase amplitude $\Psi(\phi, t)$ is in its ground state, which, for 
sufficiently high values of $Ej$, is well approximated by a Gaussian.
At $t>0$,
the height of the potential barrier decreases exponentially (here
we choose $\tau=5$ msec), and the phase amplitude spreads (Fig.
(\ref{wave},B)) untill it touches the
borders at $\pm \pi$. The wiggles in Fig.
(\ref{wave},C)) arise from the interference between
the two overlapping tails in the region around $\phi \sim \pi$, which eventually
spread over the full region Fig. (\ref{wave},D).
The period of the oscillations of the interference pattern depends on the
velocity of spreading of the phase amplitude: 
the more adiabatic the expansion, the smaller the number of oscillations.\\

\emph{Variational Approach: } We study Eq.
(\ref{dynamical_equation}) with a time-dependent variational
approach \cite{Smerzi_2000}. The time evolution of the variational
parameters is characterized by the minimization of an action with
the effective Langragian \be L(q_i, \dot{q}_i)=i \langle \Psi
\dot{\Psi} \rangle - \langle \Psi \hat{H} \Psi \rangle, \ee with
$q_i$ being the time-dependent parameters of the phase amplitude
$\Psi(\phi, q_i(t))$. This provides the familiar Lagrange
equations \be \frac{\ud }{\ud t} \frac{\partial L}{\partial
\dot{q}_i}=\frac{\partial L}{\partial q_i}. \ee We choose the
time-dependent variational phase amplitude \be \label{gaussian}
\Psi(\phi,t)=\frac{1}{(2\pi\sigma_{\phi}^2(t))^{1/4}}\exp\Big(-\frac{\phi^2}{4
\sigma_{\phi}^2(t)}+i \frac{\delta(t)}{2}\phi^2\Big), 
\ee 
with the condition that the width $\sigma_{\phi}(t)\ll 2\pi$ during the
dynamics. At the beginning of the dynamics, the phase amplitude is rather
narrow and, feeling only the quadratic part of the $\cos(\phi)$
potential, it can be approximated by
the gaussian (\ref{gaussian}). During the first stage of the dynamics, 
the phase amplitude will
follow the instantaneous ground state of the system. 
A breakdown of adiabaticity will
occur at time $t_{ad}$, which will depend on the ramping time
$\tau_R$. The Gaussian ansatz will fail when
the wave-function touches the borders 
$\phi=\pm \pi$, namely when $\sigma_{\phi}(t)\gtrsim 1$.
With the variational ansatz (\ref{gaussian}) the equation of 
motion for the width of the phase amplitude becomes:
\be \label{sigmaddot}
\ddot{\sigma}_{\phi}=\frac{1}{\sigma_{\phi}^3} - 2 \, 
\sigma_{\phi} \,\Gamma(t) \, e^{-\sigma_{\phi}^2/2}.
\ee

\begin{figure}[!h]
\begin{center}
\includegraphics[scale=1.2]{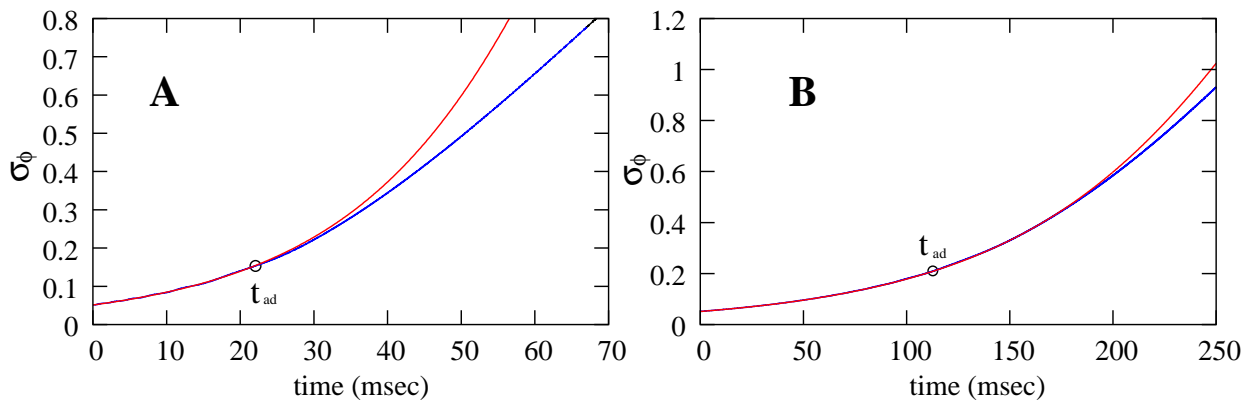}
\end{center}
\caption{\small{Comparison between different solutions of the
equation (\ref{dynamical_equation}) for two cases: A)
$\tau_R^{eff}=5$ msec, and B) $\tau_R^{eff}=20$ msec. The blue
line represents the numerical solution, and the red line
represents the variational adiabatic solution given by the
equation (\ref{sigmaad3}). We also indicate the time of breakdown
of adiabaticity as given by the eq. (\ref{t_ad}): $t_{ad}=21$ msec
for A), and $t_{ad}=111$ msec for B).}}\label{adiab}
\end{figure}

\emph{Adiabatic Variational Solution. } We can calculate the
adiabatic solution from equation (\ref{sigmaddot}) by
imposing the adiabaticity condition $\ddot{\sigma}_{\phi}=0$. 
In this limit, we have $\Gamma \gg 1$, 
and the dynamics are governed by the
particle exchange through the two wells, which keeps the phase
coherence between the two condensates. We obtain the equation 
\be \label{sigmaad3} 
\sigma_{ad}(t)=
\sqrt{4\, W\bigg(\sqrt{\frac{1}{4\sqrt{2 \Gamma(t)}}}\bigg)}=
\sqrt{4\, W\bigg(\frac{1}{8} \sqrt{\frac{E_c}{E_j(0)}} \, e^{t/2\tau}\bigg)}, 
\ee 
where $W(x)$ is the Lambert-W function \cite{MathWorld}.
In Figure (\ref{adiab}) 
we compare the adiabatic solution with the numerical one
for the values $\tau=5$ msec and $\tau=20$ msec. \\

\begin{figure}[!h]
\begin{center}
\includegraphics[scale=0.6]{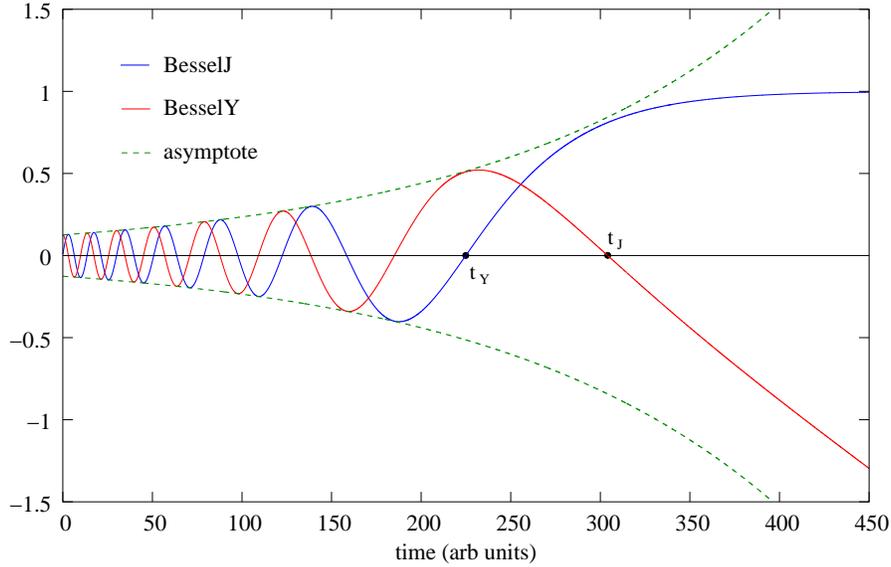}
\end{center}
\caption{\small{Behavior of the functions $J_0 (2 \tau\, e^{-t/\tau} )$ (blue line)
and $Y_0 (2 \tau \, e^{-t/\tau} )$ (red line). The dashed green line represents the 
exponential asymptotes $e^{t/4\tau}$.}}\label{graf}
\end{figure}

\emph{Linear approximation} 
We now solve equation (\ref{sigmaddot}) seeking a solution of the form
$\sigma_{\phi}(t)=\sigma_{ad}(t)+\varepsilon(t)$. We expect that
there will be a time at which the solution $\varepsilon(t)$ becomes
a significant correction to $\sigma_{ad}(t)$: this will give the criterion
for the breakdown of adiabaticity. Replacing
$\sigma_{\phi}(t)=\sigma_{ad}(t)+\varepsilon(t)$ in equation
(\ref{sigmaddot}), where $\sigma_{ad}(t)$
is the adiabatic solution (\ref{sigmaad3}), we obtain a second-order
differential equation for $\varepsilon(t)$ 
\be \label{eqvarepsilon}
\ddot{\varepsilon}(t)=-\Bigg(\frac{4+\sigma_{ad}^2(t)}{\sigma_{ad}^4(t)}\Bigg)\,
\varepsilon(t)+O\big(\varepsilon(t)^2\big), 
\ee 
where we neglected quadratic terms in $\varepsilon(t)$. 
We are interested in studying this equation for times $0\leq t \leq t_{ad}$: 
in this range we can assume $\sigma_{ad}^2(t)\ll 1$, and equivalently $E_j(t)\gg
E_c$ and $\Gamma(t)\gg 1$. 
In the regime $\sigma_{ad}\ll 1$, we approximate 
$(4+\sigma_{ad}^2(t))/\sigma_{ad}^4(t) \sim 4/\sigma_{ad}^4(t)$. In the same 
approximation we expand Eq. (\ref{sigmaad3}) in a
Taylor series for $1/\Gamma(t) \approx 0$, neglecting quadratic terms. 
Accounting for the rescaling of the time, the
equation (\ref{eqvarepsilon}) becomes 
\be \label{epsilon_solution}
\ddot{\varepsilon}(t)=-\frac{4 E_c E_j(0)}{\hbar^2} \, e^{-t/\tau} \, \varepsilon(t). 
\ee 
This is the equation of a harmonic
oscillator with a time dependent (exponentially-decreasing)
frequency. It is possible to find the solution of this equation in
terms of Bessel functions: \be \label{epsilon} \varepsilon(t)=C\,
J_0 \Bigg(2 \sqrt{\frac{4 E_c E_j(0)}{\hbar^2}}\,\tau \,
e^{-t/2\tau}  \Bigg)+ D\, Y_0 \Bigg(2 \sqrt{\frac{4 E_c
E_j(0)}{\hbar^2}}\,\tau\, e^{-t/2\tau}  \Bigg), \ee where $J_0$ is
the zero-order Bessel function of the first kind and $Y_0$ is the
zero-order Bessel function of the second kind \cite{Abramovitz},
while $C$ and $D$ are constants which depend on the initial conditions
$\varepsilon (0)$, and $\dot{\varepsilon}(0)$. Choosing as initial
condition $\varepsilon (0)=0$, and
$\dot{\varepsilon}(0) \neq 0$, we have
\be \label{C}
C=-\frac{\dot{\varepsilon}(0)}{\sqrt{A}} \,
\Bigg(\frac{Y_0(2\sqrt{A}\tau)}{Y_1(2\sqrt{A}\tau)J_0(2\sqrt{A}\tau)-Y_0(2\sqrt{A}\tau)J_1(2\sqrt{A}\tau)}\Bigg),
\ee \be \label{D} D=\frac{\dot{\varepsilon}(0)}{\sqrt{A}} \,
\Bigg(\frac{J_0(2\sqrt{A}\tau)}{Y_1(2\sqrt{A}\tau)J_0(2\sqrt{A}\tau)-Y_0(2\sqrt{A}\tau)J_1(2\sqrt{A}\tau)}\Bigg),
\ee where $A=4 E_c E_j(0)/\hbar^2$. 
Figure (\ref{graf}) presents a plot of the
functions $J_0 \big(2 \sqrt{A}\,\tau\, e^{-t/2\tau}  \big)$
and $Y_0 \big(2 \sqrt{A}\,\tau\, e^{-t/2\tau}  \big)$ for $A=1$ 
and for an arbitrary time $\tau$.\\
\\*
Initially, the solutions are oscillating with a 
$\pi/2$ phase difference: \be J_0\big(2
\sqrt{A}\,\tau\, e^{-t/2\tau}  \big)\approx \sqrt{\frac{1}{\pi
\sqrt{A} \tau}}\, e^{t/4\tau}\cos\Big(2 \sqrt{A} \tau
e^{-t/2\tau}-\frac{\pi}{2} \Big), \ee \be Y_0\big(2
\sqrt{A}\,\tau\, e^{-t/2\tau}  \big)\approx \sqrt{\frac{1}{\pi
\sqrt{A} \tau}}\, e^{t/4\tau}\sin\Big(2 \sqrt{A} \tau
e^{-t/2\tau}-\frac{\pi}{2} \Big). \ee 
The amplitude of the oscillations
increase exponentially as $e^{-t/4\tau}$. For large
times ($t \gtrsim 4 \tau$) the functions approach their asymptotic
behavior: $J_0$ tends to a constant value \be J_0\big(2
\sqrt{A}\,\tau\, e^{-t/2\tau}  \big) \to 1 \ee independently of
$A$ and $\tau$, and $Y_0$ diverges linearly in time \be Y_0\big(2
\sqrt{A}\,\tau\, e^{-t/2\tau}  \big) \to -\frac{2}{\pi} \ln\big(2
\sqrt{A} \tau \big) \, \frac{t}{2\tau}. \ee The times $t_J$ and
$t_Y$ in the figure (\ref{graf}) indicate the last oscillation of
the Bessel functions $Y_0$ and $J_0$. They are defined by the
equations \be \label{t_Y} 2 \sqrt{A} \tau e^{-t_Y/2\tau}=0.8935,
\ee \be \label{t_J} 2 \sqrt{A} \tau e^{-t_J/2\tau}=2.4048. \ee 
We interpret this divergence at $t > 4 \tau$ as the signature of 
breakdown of adiabaticity.
From equations (\ref{t_Y}) and (\ref{t_J}) it is natural to
define the time of breakdown of adiabaticity $t_{ad}$ by the
relation 
\be 
\label{t_ad} 2 \sqrt{A} \tau e^{-t_{ad}/2\tau}=c, 
\ee
where $c$ is an arbitrary number $\sim 1$. 
Since $A=4 E_c E_j(0)/\hbar^2$,
$E_J(t)=E_J(0)e^{-t/\tau}$, and introducing the Josephson
oscillation frequency $\omega_j(t)=\sqrt{E_c E_j(t)}/\hbar$, we
can rewrite Eq. (\ref{t_ad}) as \be \label{t_ad2}
\frac{1}{\omega_j(t_{ad})}=\frac{4}{c}\, \tau, \ee which gives
\be \label{t_ad3} t_{ad}=2 \tau \ln\Big(\omega_j(0)\, \frac{4}{c}
\, \tau \Big). 
\ee 
Eq. (\ref{t_ad2}) coincides with the definition of
breakdown of adiabaticity suggested with a different heuristic
argument by Javanainen in
\cite{Javanainen_1999} with $\alpha=4/c$. In his work, Javanainen
estimates numerically the parameter $\alpha=2\pi$. This value is
in reasonable agreement with our analytical calculations. We have
$c=2/\pi=0.63$, which is in good agreement with our estimations.
The more important physical result suggested by equation
(\ref{t_ad2}) is the fact that the constant $c$ does not depend of
the time scale $\tau$. This has a physical meaning
\cite{Legget_Sols_1998}. Initially the system is in its ground
state, and we suppose $E_j(0) \gg E_c$; the phase
dispersion is $\sigma_{\phi}(0)\ll 1$, and the system feels only
the quadratic part of the $\cos(\phi)$ potential. By ramping the
two wells, the quantity $E_j(t)$ decreases with time scale $\tau$.
As long as $\omega_j(t) \gg 1/\tau$, the system adjusts itself in
such a way that it is always in the ground state, and the change of
the potential is adiabatic. Over the time $t_{ad}$ defined by the
equation (\ref{t_ad2}), the frequency $\omega_j(t)$ becomes so
small that it is impossible for the system to adjust in the
ground state following the decreasing of the tunneling rate.
The evolution of the system is no longer adiabatic.\\

Figure (\ref{adiab}) shows a comparison between the numerical
solution (blue line) of the dynamical equation
(\ref{dynamical_equation}) and the variational adiabatic solution
(red line) given by the equation (\ref{sigmaad3}). The figures
(\ref{adiab}A, B) refer to the cases $\tau=5$ msec and
$\tau_R=20$ msec, respectively. In these figures  the time
$t_{ad}$ is defined by equation (\ref{t_ad}) with $c=1$.
\\*

\begin{figure}[!h]
\begin{center}
\includegraphics[scale=0.6]{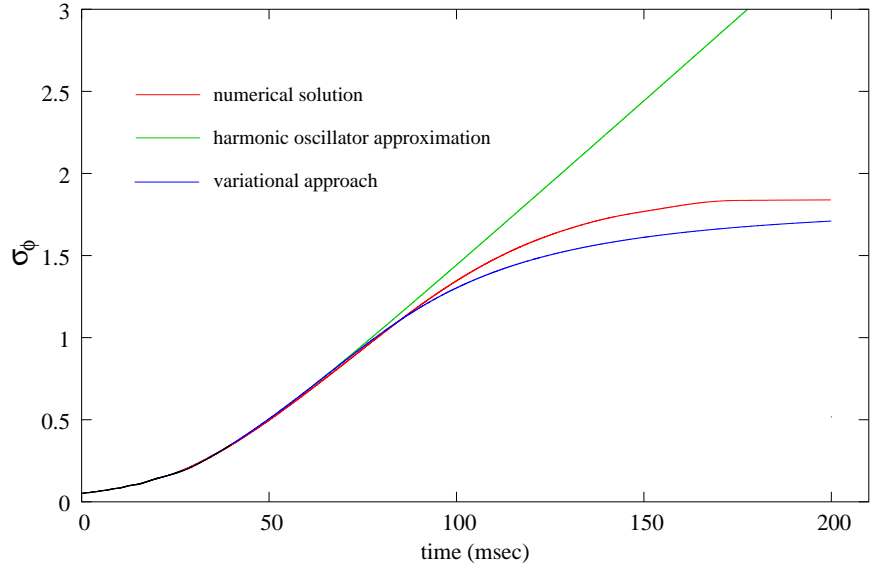}
\end{center}
\caption{\small{Comparison between the numerical solution of
equation (\ref{dynamical_equation}) (red line), the variational
approach obtained by solving the equation (\ref{sigmaddot}) (blue
line), and the harmonic oscillator approximation obtained by
substituting $\cos \phi \to -\phi^2/2$ in Eq.
(\ref{dynamical_equation})}}.\label{tau5}
\end{figure}

\begin{figure}[!h]
\begin{center}
\includegraphics[scale=0.5]{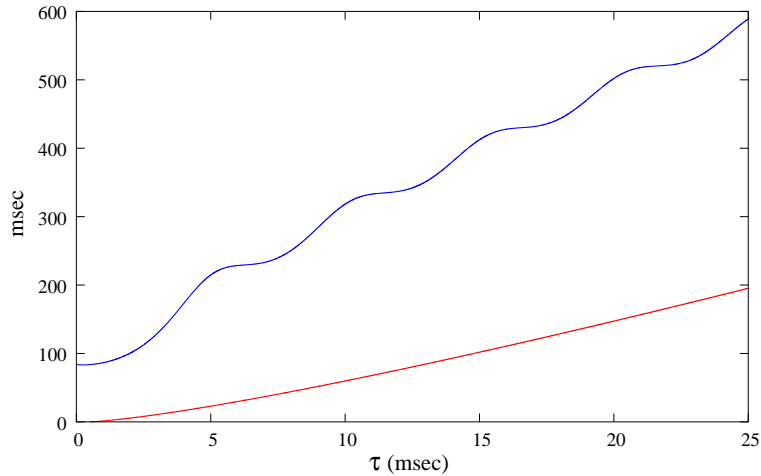}
\end{center}
\caption{\small{Dependences of $t_D$ (blue line) and $t_{ad}$ (red
line) as the functions of $\tau$. The dephasing time is given by
Eq. (\ref{t_D}), the time of breakdown of adiabaticity is given by
Eq. (\ref{t_ad}).}}\label{time}
\end{figure}

\emph{Dephasing Time: } We define the dephasing time $t_D$ as
the time needed for the phase dispersion to become of order 1.
Roughly, at $t=t_D$ the phase amplitude $\Psi(\phi,t)$ reaches the borders
$\phi=\pm \pi$, and we lose every information about the phase. It might
be useful to recall that in a single experiment a well defined phase will
actually be measured. However, in different experiments, repeated in identical
conditions, the measured phases would differ, with a mean square
fluctuation $\sigma_{\phi}$. The main result of this
section will be the analytical estimate of $t_D$. We first notice
that (comparing
the red and blue lines in the figure (\ref{tau5})) 
we can  approximate well the numerically exact phase amplitude with a
gaussian even for $t>t_{ad}$. We can therefore expect that
the gaussian variational ansatz can give a good estimate of $t_D$.
Thus, we approximate the QPM
$\hat{H}_{\phi}$ with the Hamiltonian of a harmonic oscillator with
a time dependent frequency: 
\be
H_{\phi}=-\frac{E_c}{2\hbar}\frac{\partial^2}{\partial
\phi^2}-\frac{E_j(0)\, e^{-t/\tau}}{2\hbar}\, \phi^2. 
\label{tdho}
\ee 
The quantum evolution of the system governed by Eq.(\ref{tdho})
can be calculated exactly in the Wigner phase space.
We first need to calculate the solution of the classical equation of
motion
\be
\ddot{\phi}(t)=-\frac{E_c \, E_j(0)}{\hbar^2}\,e^{-t/\tau}\,
\phi(t), \ee 
which can be expressed in terms of 
Bessel functions of the first and the second kind. With
$A=E_c \, E_j(0)/\hbar^2$, we have (similarly with equation
(\ref{epsilon_solution}) and (\ref{epsilon})): \be
\label{solution} \phi(t)=C \, J_0(2\sqrt{A}\tau e^{-t/2\tau})+D \,
Y_0(2\sqrt{A}\tau e^{-t/2\tau}), \ee where $C$ and $D$ depend of
the initial conditions $\phi_0\equiv \phi(0)$ and
$\dot{\phi}_0\equiv\dot{\phi}(0)$: \be
C=\frac{\phi_0}{J_0(2\sqrt{A}\tau)}-\frac{Y_0(2\sqrt{A}\tau)}{J_0(2\sqrt{A}\tau)}
\frac{\frac{\dot{\phi}_0}{\sqrt{A}}J_0(2\sqrt{A}\tau)-\phi_0J_1(2\sqrt{A}\tau)}
{Y_1(2\sqrt{A}\tau)J_0(2\sqrt{A}\tau)-Y_0(2\sqrt{A}\tau)J_1(2\sqrt{A}\tau)},
\ee \be D=
\frac{\frac{\dot{\phi}_0}{\sqrt{A}}J_0(2\sqrt{A}\tau)-\phi_0J_1(2\sqrt{A}\tau)}
{Y_1(2\sqrt{A}\tau)J_0(2\sqrt{A}\tau)-Y_0(2\sqrt{A}\tau)J_1(2\sqrt{A}\tau)}.
\ee 
We can now calculate the phase dispersion 
$\sigma_{\phi}^2(t)=\langle \, \phi(t, \phi_0, \dot{\phi}_0)^2\,
\rangle-\langle \, \phi(t, \phi_0, \dot{\phi}_0)\, \rangle^2$,
where the brackets indicate the integration over the initial
conditions $\phi_0$ and $\dot{\phi}_0$ averaged with the Wigner
transform $P(\phi_0, \dot{\phi}_0, t)$. For an harmonic oscillator
the Wigner transform \cite{Hillery_1984} becomes
\begin{eqnarray} \label{P}
P(\phi_0, \dot{\phi}_0, t)&=&
\frac{1}{\pi}\frac{\hbar}{E_c}\int_{-\infty}^{+\infty} \ud \xi \,
\frac{e^{-\frac{(\phi_0-\xi)^2}{4\sigma_{\phi_0}^2}}}{(2\pi\sigma_{\phi_0}^2)^{\frac{1}{4}}}\,
\frac{e^{-\frac{(\phi_0+\xi)^2}{4\sigma_{\phi_0}^2}}}{(2\pi\sigma_{\phi_0}^2)^{\frac{1}{4}}}\,
e^{2i\frac{\hbar}{E_c}\xi\dot{\phi}_0} \\
&=&
\frac{e^{-\frac{\phi_0^2}{2\sigma_{\phi_0}^2}}}{(2\pi\sigma_{\phi_0}^2)^{\frac{1}{4}}}\,
\frac{e^{-\frac{{\dot{\phi}_0}^2}{2\sigma_{{\dot{\phi}}_0}^2}}}{(2\pi\sigma_{{\dot{\phi}}_0}^2)^{\frac{1}{4}}}, \\
\nonumber
\end{eqnarray}
where $\sigma_{\dot{\phi}_0}=E_c/(2 \hbar\sigma_{\phi})$ is the width of the distribution 
$P(\phi_0, \dot{\phi}_0, t)$ in the variable $\dot{\phi}_0$ canonically conjugated to $\phi_0$. 
Notice that, in equation (\ref{P}), 
we have integrated between $\pm \infty$, and not
between $\pm \pi$, which is consistent with 
the condition $\sigma_{\phi_0},\sigma_{\dot{\phi}_0}\ll 2\pi$.
Since the Wigner transform (\ref{P}) is symmetric in $\phi_0$ and
$\dot{\phi}_0$, we have that $\langle \, \phi(t, \phi_0,
\dot{\phi}_0)\, \rangle=0$, and
\begin{eqnarray} \label{solution2}
\sigma_{\phi}(t)^2 &=& \langle \, \phi(t, \phi_0, \dot{\phi}_0)^2\, \rangle \nonumber\\
&=& \int_{-\infty}^{+\infty} \ud \phi_0 \ud \dot{\phi}_0 \, \Big(C(\phi_0, \dot{\phi}_0) 
\, J_0(2\sqrt{A}\tau e^{-t/2\tau})+ D(\phi_0, \dot{\phi}_0) \, Y_0(2\sqrt{A}\tau
e^{-t/2\tau})\Big)^2 \, P(\phi_0, \dot{\phi}_0, t) \nonumber\\
&=& \Big(
C^* \, J_0\big(2\sqrt{A}\tau e^{-t/2\tau}\big)+
D^* \, Y_0\big(2\sqrt{A}\tau e^{-t/2\tau}\big) \Big)^2, \\
\nonumber
\end{eqnarray}
where $C^* (D^*)=\int_{-\infty}^{+\infty} \ud \phi_0 \ud \dot{\phi}_0 \,
C(\phi_0, \dot{\phi}_0)~ (D(\phi_0, \dot{\phi}_0))\,
P(\phi_0, \dot{\phi}_0, t)$. If we define $K(2\sqrt{A}\tau)\equiv
Y_1(2\sqrt{A}\tau)J_0(2\sqrt{A}\tau)-Y_0(2\sqrt{A}\tau)J_1(2\sqrt{A}\tau)$,
we can write \be
(C^*)^2=\frac{\sigma_{\phi}^2(0)}{J_0^2(2\sqrt{A}\tau)}
\Bigg[1+\frac{Y_0(2\sqrt{A}\tau)J_1(2\sqrt{A}\tau)}{K(2\sqrt{A}\tau)}\Bigg]^2
+\frac{\sigma_{\dot{\phi}}^2(0)}{A}\frac{Y_0^2(2\sqrt{A}\tau)}{K^2(2\sqrt{A}\tau)},
\ee \be
(D^*)^2=\frac{\sigma_{\dot{\phi}}^2(0)}{A}\frac{J_0^2(2\sqrt{A}\tau)}{K^2(2\sqrt{A}\tau)}
+\sigma_{\phi}^2(0)
\frac{J_1^2(2\sqrt{A}\tau)}{K^2(2\sqrt{A}\tau)}. \ee 
Notice that in the limit $\tau \to 0$, we recover the free
evolution dynamics as given by equation (\ref{free}). As shown by 
figure (\ref{graf}), the harmonic oscillator approximation
remain valid even for $t>t_{ad}$. The dephasing of the phase amplitude
occurs at $t_D > \tau$. We can therefore use the asymptotic expansion $t\gg
\tau$ for the Bessel functions $J_O(2\sqrt{A}\tau e^{-t/2\tau})\to
1$ and $Y_O(2\sqrt{A}\tau e^{-t/2\tau}) \to
\frac{2}{\pi}\ln(\sqrt{A}\tau)-\frac{t}{\pi\tau}$. We can rewrite
eq. (\ref{solution2}) as \be
\sigma_{\phi}(t)^2=\Bigg(C^*+D^*\Big(\frac{2}{\pi}\ln(\sqrt{A}\tau)-\frac{t}{\pi\tau}\Big)
\Bigg)^2 . \ee The dephasing time $t_D$ is defined by the relation
$\sigma_{\phi}(t_D)=1$, from which we obtain \be \label{t_D}
t_D=2\tau\ln(\sqrt{A}\tau)+\pi\tau\Big(\frac{1+C^*}{D^*}\Big). \ee
This is the central result of our paper. Figure (\ref{time}) 
presents the dependences $t_D$ and $t_{ad}$ as a function of $\tau$.\\

\begin{figure}[!h]
\begin{center}
\includegraphics[scale=0.5]{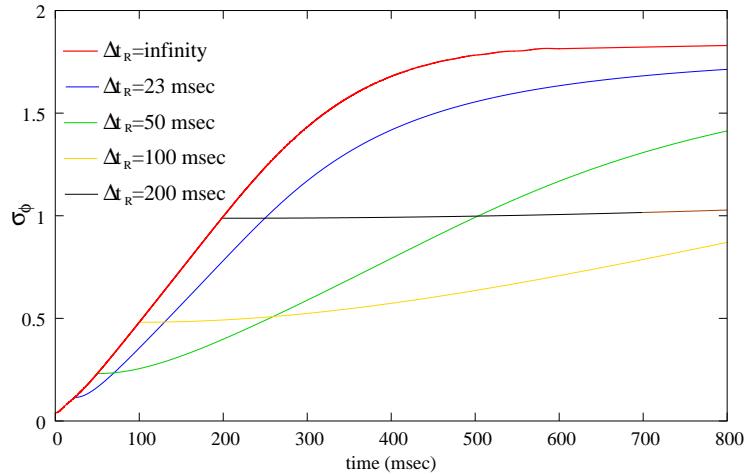}
\end{center}
\caption{\small{Evolution of $\sigma_{\phi}(t)$ for different
values of $\Delta t_R$. The red line represents the case $\Delta
t_R=\infty$ corresponding to an unlimited ramping the two potential wells.
The other lines refer to different choices of finite $\Delta t_R$.
We approximate the evolution of $\sigma_{\phi}(t)$ after the
ramping of the wells by the free evolution given by equation
(\ref{free}), and we take into account the boundaries $\pm \pi$ through
the equation (\ref{sigmat}).}}\label{freeplot}
\end{figure}

\emph{Holding time: } In \cite{Shin_2004} the
two potential wells are separated with a ramping
time $\Delta t_R$. At the end of the ramping, the condensates are held
in the trap for a time $\Delta t_{hold}$. We can study the phase
dispersion during this time by using the variational approximation
(\ref{sigmaddot}) with a constant parameter $\Gamma$. If we have
$\Delta t_R = \beta \tau$ where the $\beta \gtrsim 1$, we can
neglect the potential energy term in Eq. (\ref{sigmaddot}), which becomes
\be \label{sigma_free}
\ddot{\sigma}_{\phi}=\frac{1}{\sigma_{\phi}^3}. 
\ee
The phase-width evolves freely in time. Accounting for
the rescaling of the time, the solution of this equation is given
by 
\be \label{free}
\sigma_{\phi}^2(t)=
\sigma_{\phi}^2(\Delta t_R)+
\frac{E_c^2}{4 \, \hbar^2 \, \sigma_{\phi}^2(\Delta t_R)}\, (t-\Delta t_R)^2, 
\ee 
where $\sigma_{\phi}^2(\Delta t_R)$ is the gaussian width at
$t=\Delta t_R$ (end of the ramping). 
For $t \lesssim t_{\Delta t_R}+2\sigma_{\phi}^2(\Delta t_R)\hbar/E_c$, we have that
$\sigma_{\phi}(t)$ is almost constant, while for $t \gg
\Delta t_R+2\sigma_{\phi}^2(\Delta t_R)\hbar/E_c$ the width
$\sigma_{\phi}(t)$ is a linear function of time. 
In figure (\ref{freeplot}) we present the
evolution of $\sigma_{\phi}(t)$ for different values of $\Delta
t_R$. In this figure $\tau=5$ msec, $E_c=0.001\times\hbar$ kHz,
and $E_c=100\times\hbar$ kHz. To take into account the
boundaries $\phi=\pm \pi$, we compare the numerical solution with
the quantity \be \label{sigmat}
\tilde{\sigma}_{\phi}^2(t)=\int_{-\pi}^{+\pi} \ud \phi \, |\psi(\phi)|^2
\, \phi^2, \ee
where $|\psi(\phi)|^2 \sim \exp(-\phi^2/2\sigma_{\phi}^2(t))$ and $\sigma_{\phi}^2(t)$ is given by Eq. (\ref{free}).\\
In the model of Legget and Sols \cite{Legget_Sols_1998,Leggett_1991,Zapata_1998} 
the free evolution takes
place with the breakdown of adiabaticity. This model assumes that
$\Delta t_R=t_{ad}$. We can check this model by looking at Fig.
(\ref{free}): the red line represents the numerical solution of
equation (\ref{dynamical_equation}), for $\tau=5$ msec, the
blue line represents the free expansion occurring at the breakdown
of adiabaticity $t_{ad}$ as defined by eq. (\ref{t_ad}). We can
see that the model of Legget and Sols gives a reasonable
estimation of the phase spreading, however it over-estimates the
dephasing time.\\

\emph{Repeated Interference Experiments: } It has been recently shown by 
Castin and Dalibard
\cite{Castin_1997} and Javanainen and Yoo \cite{Javanainen_1996}
that the phase of a condensate is established by
measurement. Two BECs, initially in
number "Foch" states, will interfere and have a
definite relative phase. A different phase, however, will be measured in
different experiments, so that averaging over the ensemble,
no interference is observed.
To illustrate the effect of dephasing, we present here a
simple qualitative analysis. We consider the interference between
two independent (not overlapping) condensates in the Thomas Fermi (TF)
approximation. Initially the two condensates are in equilibrium,
the condensate 1 being centered at $x=-d/2$, and the condensate 2 at
$x=d/2$. We assume that at $t=0$ the order parameter is
described by the linear combination \be \label{ansatzinterf}
\hat{\Psi}(\vec{r})=\hat{\Psi}_1(\vec{r})+e^{i
\Phi}\hat{\Psi}_2(\vec{r}), \ee where $\Psi_1(\vec{r})$ and
$\Psi_2(\vec{r})$ are the equilibrium wave functions (order
parameters) of the two condensates, respectively, which we assume are well
separated in space: 
\be 
\label{overlap}
\Psi_1(\vec{r})\Psi_2(\vec{r})\approx 0. \ee
At $t=0$ the trap is switched off. We
neglect the interaction between the two condensates but we
account for the atom-atom interaction in each single
condensate, which is important during the free expansion. 
The total density $\rho=|\Psi|^2$ of the overlapping 
condensates exhibits modulation of the form
\be \rho(\vec{r},t, \phi)=\rho_1(\vec{r},t)+\rho_2(\vec{r},t)+ 2
\sqrt{\rho_1(\vec{r},t)\rho_2(\vec{r},t)} \, \cos
\Big(S_1(\vec{r},t)-S_2(\vec{r},t)+\phi \Big), \ee where
$\Psi_{1,2}(\vec{r},t)=\sqrt{\rho_{1,2}(\vec{r},t)}\,
e^{S_{1,2}(\vec{r},t)}$. By recalling the quadratic behavior of
the phase in the TF approximation, we obtain asymptotically 
\be
S_1(\vec{r},t)-S_2(\vec{r},t)=\frac{md}{\hbar t} \, x. 
\ee
\begin{figure}
\includegraphics[width=3.5in]{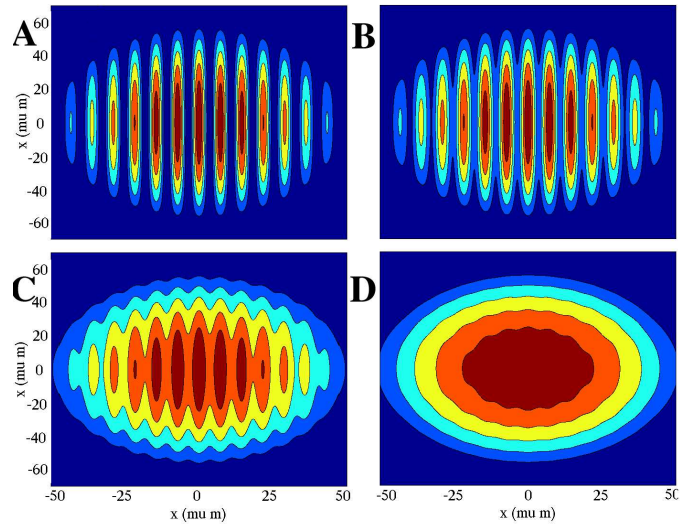}
\caption{\small{Contour plot of the interference fringes
calculated in the TF approximation. The density is averaged over
several experiments, according to Eq. (\ref{average}). The quantum
dynamics determines a phase uncertainty $\sigma$: A) $\sigma=0$,
B) $\sigma=1$, C) $\sigma=2$, D) $\sigma=3$.}} \label{TF}
\end{figure}
In a single experiment, the interference pattern is characterized
by the straight line fringes which are orthogonal to the $x$-axis
(radial axis of the two cigar shaped parallel condensates). We
obtain fringes perpendicular to the $x$-axis with spacing $h
t/md$ between two consecutive fringes. To experimentally test the
quantum phase dynamics it would be necessary to average
over several identical interferometric realizations. The relative
phase will be chosen randomly with a gaussian distribution of
width $\sigma$. The ensemble averaged density is therefore
$\rho(\vec{r},t)= \int_{-\pi}^{+\pi}\ud \phi \,
\rho(\vec{r},t;\phi) |\Psi(\phi,t)|^2 ={1 \over
2}[\rho_1(\vec{r},t)+\rho_2(\vec{r},t)+\rho_{int}(\vec{r},t)]$,
with $\rho_{1,2}(\vec{r},t)  \equiv |\psi_{1,2}(\vec{r},t)|$ the
densities of each released condensate, and
\begin{eqnarray} \label{average}
\rho_{int}(\vec{r},t)& = & 2 {\sqrt{\rho_1(\vec{r},t)\rho_2(\vec{r},t)}}
\int_{-\pi}^{+\pi}\ud \phi
\cos\Big( \frac{md}{\hbar t} \, x +\phi\Big)\frac{e^{-\frac{\phi^2}{2\sigma^2}}}{\sqrt{2 \pi \sigma^2}} \\
& = & 2 {\sqrt{\rho_1(\vec{r},t)\rho_2(\vec{r},t)}}  \cos\Big(\frac{md}{\hbar t} \, x\Big)e^{-\sigma^2/2}. \\
\nonumber
\end{eqnarray}
\\*

\emph{Conclusions: }
We have studied the dephasing and adiabaticity time scales in the splitting of a single 
Bose Einstein Condensate. We have focused on the quantum behavior of 
the system within a two-mode model,
solving a Schr\"odinger-like equation in the phase variable within a variational gaussian ansatz. 
We have obtained the exact behavior in the adiabatic and free expansion regimes, and we have obtained   
analytical estimations for the dephasing time and the time when the dynamics depart from the 
adiabatic evolution. 
We have compared our analytical results with both full numerical 
solutions of the two-mode Hamiltonian, and the existing literature. 
Finally, we have shown how the spreading of phase distribution can affect the visibility 
of interference fringes in an ensemble of repeated experiments. \\ 
\\*  
This work  was supported by the Department of Energy under the
contract W-7405-ENG-36 and DOE Office of Basic Energy Sciences.

\end{document}